\documentclass{pasa}
\usepackage{graphicx,url,amssymb,amsmath,rotating,color,units,wasysym,epsfig,multirow,epstopdf,subcaption,xcolor}
\usepackage{subcaption}
\usepackage{tikz}
\usetikzlibrary{trees,positioning,shapes,shadows,arrows}
\tikzset{
  basic/.style  = {draw, text width=5cm, rectangle},
  root/.style   = {basic, rounded corners=3pt, thin, align=center, fill=white},
  level-2/.style = {basic, rounded corners=3pt, thin, align=center, fill=white, text width=2.2cm},
  level-3/.style = {basic, rounded corners=3pt, thin, align=center, fill=white, text width=1.7cm}
}
\usepackage{hyperref}

\title[When models fail]{When models fail: an introduction to posterior predictive checks and model misspecification in gravitational-wave astronomy}

\author{
Isobel Romero-Shaw$^{1,2}$\thanks{isobel.romero-shaw@monash.edu},
Eric Thrane$^{1,2}$\thanks{eric.thrane@monash.edu}\, and
Paul D. Lasky$^{1,2}$\thanks{paul.lasky@monash.edu}
\affil{$^1$Monash Astrophysics, School of Physics and Astronomy, Monash University, VIC 3800, Australia}%
\affil{$^2$OzGrav: The ARC Centre of Excellence for Gravitational-Wave Discovery, Clayton, VIC 3800, Australia}
}%

\jid{PASA}
\doi{10.1017/pas.\the\year.xxx}
\jyear{\the\year}

\usepackage{aas_macros}
\usepackage{hyperref} 
\hypersetup{colorlinks,citecolor=blue,linkcolor=blue,urlcolor=blue}

\begin{document}

\begin{frontmatter}
\maketitle

\begin{abstract}
Bayesian inference is a powerful tool in gravitational-wave astronomy.
It enables us to deduce the properties of merging compact-object binaries and to determine how these mergers are distributed as a population according to mass, spin, and redshift.
As key results are increasingly derived using Bayesian inference, there is increasing scrutiny on Bayesian methods.
In this review, we discuss the phenomenon of \textit{model misspecification}, in which results obtained with Bayesian inference are misleading because of deficiencies in the assumed model(s).
Such deficiencies can impede our inferences of the true parameters describing physical systems.
They can also reduce our ability to distinguish the ``best fitting'' model: it can be misleading to say that Model~A is preferred over Model~B if both models are manifestly poor descriptions of reality.
Broadly speaking, there are two ways in which models fail:
models that fail to adequately describe the data (either the signal or the noise) have misspecified likelihoods.
Population models---designed, for example, to describe the distribution of black hole masses---may fail to adequately describe the true population due to a misspecified prior.
We recommend tests and checks that are useful for spotting misspecified models using examples inspired by gravitational-wave astronomy.
We include companion python notebooks to illustrate essential concepts.
\end{abstract}

\end{frontmatter}

\section{Introduction}
Bayesian inference and parameter estimation are the cornerstones of gravitational-wave astronomy.
The Bayesian framework is used to derive posterior distributions for parameters such as the masses and spins of merging pairs of neutron stars and black holes.
Using Bayesian inference, we obtain values for the marginal likelihood (also known as the evidence), which are used for model selection---for example, to compare alternative theories of gravity with general relativity. Recent discoveries that rely heavily on Bayesian inference include the extreme-mass-ratio binary GW190814~\citep{GW190814}, containing either the least massive known black hole or the most massive known neutron star; the intermediate-mass black hole event GW190521 \citep{GW190521}; multiple binaries containing a black hole paired with a neutron star \citep{NSBH,GWTC-3}; and currently the most massive binary black hole candidate, GW190426\_190642 \citep{GWTC-2-final}.

A second layer of Bayesian analysis is built upon this foundation to study the population properties of merging binaries.
Hierarchical models are used to estimate population hyper-parameters describing how sources of gravitational waves are distributed according to mass, spin, redshift, and so on. This has been key in, for example, the discovery of features in the distribution of binary black hole masses \citep{o3a_pop, GWTC-3-pop}.

Results obtained with Bayesian inference are only as reliable as the underlying model.
The compact-object binary parameters reported in gravitational-wave transient catalogues are derived using models that describe physical systems; only if these models are sufficient descriptors of the true system can these results be meaningful.
Bayesian inference can tell us that one model is a better explanation for the data than another.
For example, Bayesian techniques have been used to suggest that intermediate-mass black hole merger GW190521 shows signs of non-zero orbital eccentricity \citep{GW190521_formation}.
However, Bayesian inference \textit{on its own} does not tell us if either the quasi-circular or eccentric gravitational waveforms considered provide an adequate fit to the GW190521 data.
Similarly, Bayesian inference has been used to suggest that the distribution of primary black hole mass is better fit by a broken power-law distribution than a power-law distribution with no break \citep{o3a_pop}.
However, Bayesian inference \textit{on its own} does not tell us if either of these models is adequate to describe the observed distribution of primary black hole masses.

As the gravitational-wave catalogue grows and gravitational-wave detector sensitivity improves, we begin to see more events that push the boundaries of our understanding of the Universe.
This makes it ever more important to test the validity of our models.
A signal model that is valid for systems with mass ratios $q \geq 0.125$ may be invalid for a mass ratio of $q = 0.001$, and a detector noise model adequate for an event with signal-to-noise ratio $\text{SNR}=30$ may be inadequate for an $\text{SNR}=100$ signal.
Additionally, as the number of gravitational-wave signal detections grows, the resolving power of the combined dataset increases.
This makes it ever more important to test the validity of population models.
A population model for the distribution of binary black hole redshifts that works reasonably well for a dozen events may be unsuitable for a catalogue with hundreds of events.

In this Article, we describe different ways in which models can fail and lay out commonly-used tests that can be carried out to reveal these failures, often beginning with visualisation.
For a broader discussion of how visualisation can assist in solving Bayesian inference problems, see \cite{Gabry:2017:visualisation}.
We discuss workarounds for misspecified models, including model redesign and data coarsening \citep{Miller2019,Thomas2019}.
For an idea of how Bayesian inference problems may be solved through a careful workflow and iterative model redesign of the prior and likelihood, see both \citet{Betancourt-workflow-website} and \citet{Gelman:2020:Workflow}.
While we cast many examples in the language of gravitational-wave astronomy, we endeavour to use sufficiently general language so that this review is useful to a broad audience.
For additional resources, we refer the reader to Chapters 6-7 of \cite{Gelman}, respectively devoted to ``Model checking'' and ``Evaluating, comparing, and expanding models.''
See also ``Model Checking and Sensitivity Analysis'' in \cite{Springer4}.

Almost all of the Subsections in this review follow the same formula.
After introducing a concept, we provide a bullet list of \textit{recommended tests}.
This list is followed by a \textit{demonstration}, which illustrates the tests with simple examples.
This layout is designed to help researchers scan the Article to quickly find the misspecification tests they are looking for.
Our recommended tests do not constitute an exhaustive list of the ways in which one may test for misspecification.
In addition, an analysis that passes all of these tests may still suffer from misspecification.
There is no silver bullet to detect all forms of misspecification!
Nonetheless, the tests recommended here provide a useful starting point for checking the suitability of models.
All of the demonstration code is available in \texttt{jupyter} notebook form here: \href{https://tinyurl.com/bf4n9vw5}{\textcolor{blue}{tinyurl.com/bf4n9vw5}}.
There is a dedicated notebook for each Section.

Broadly speaking, two different kinds of models are required to do an inference calculation.
Every such calculation requires a model for the distribution of the data---the likelihood function---and a model for the distribution of the parameters---the prior.
The remainder of this Article is organised as follows.
In Section~\ref{visualisation}, we discuss the importance of data visualisation.
In Section~\ref{data}, we describe misspecification of the data: misspecified likelihood functions.
In Section~\ref{priors}, we describe misspecification of population models: misspecified priors.
In Section~\ref{other}, we describe how apparent outliers may or may not be signs of model misspecification.
We provide closing thoughts in Section~\ref{conclusions}.


\section{Preface: the importance of visualisation}\label{visualisation}
Below we describe many goodness-of-fit tests that can be used to determine the suitability of different models.
However, even the most carefully crafted tests are no replacement for sanity checks with visualisation.
Plotting data, we can sometimes see obvious model failures that we might not have thought to check \textit{a priori}.
The importance of visualisation is dramatically illustrated by Anscombe's Quartet \citep{Anscombe}: four 11-point datasets with noticeably different trends that nonetheless have near-identical simple descriptive statistics.\footnote{For a more recent and creative example, see the Datasaurus Dozen \citep{Datasaurus}.}

\begin{figure}
    \centering
    \includegraphics[width=0.5\textwidth]{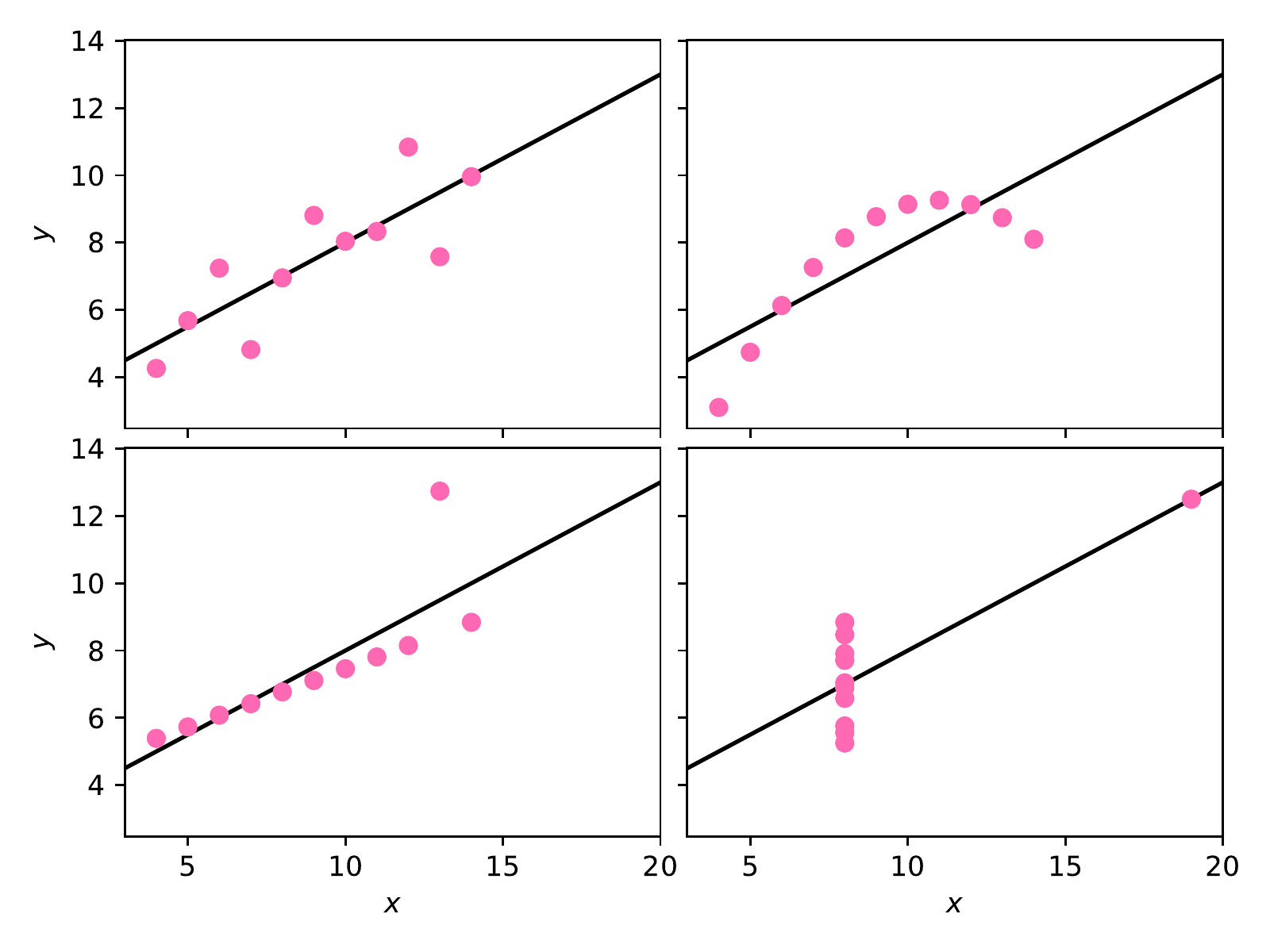}
    \caption{Anscombe's Quartet: a demonstration of the importance of data visualisation. While these datasets appear very different when plotted, they have identical summary statistics: mean $\bar{x} = 9, \bar{y} = 7.50$, sample variance $s^2_{x} = 9, s^2_{y} = 4.125 \pm 0.003$, $x-y$ correlation coefficient $0.816$, linear regression line $y_R = 3.00 + 0.500 x_R$, and linear regression coefficient of determination $R^2 = 0.67$. An \href{https://colab.research.google.com/drive/1tVa4K8mtXbKcxM1dEnMy8rRkvi-vZnTB?usp=sharing}{\textit{Anscombe's Quartet}} notebook is provided to demonstrate the calculation of these summary statistics for these datasets.}
    \label{fig:anscombe}
\end{figure}
The four datasets that comprise Anscombe's Quartet are plotted in Figure \ref{fig:anscombe}.
By studying these graphs, one can begin to diagnose anomalies: for example, two of the data sets each contain a single outlier that skews the correlation coefficient (lower left) or implies the existence of a relationship that is not supported by the rest of the data (lower right).
We can also see that the dataset in the top right would be better-specified by a non-linear relationship between $x$ and $y$.
The Quartet is a cautionary tale to those who wish to establish the ``goodness'' of their model: if one's model does not well-specify one's data, then the calculated ``goodness'' metric is not to be trusted.
The starting point for any exploration of misspecification, therefore, should be to visually compare the model and the data.

\section{Likelihood misspecification}\label{data}
\subsection{Basics}

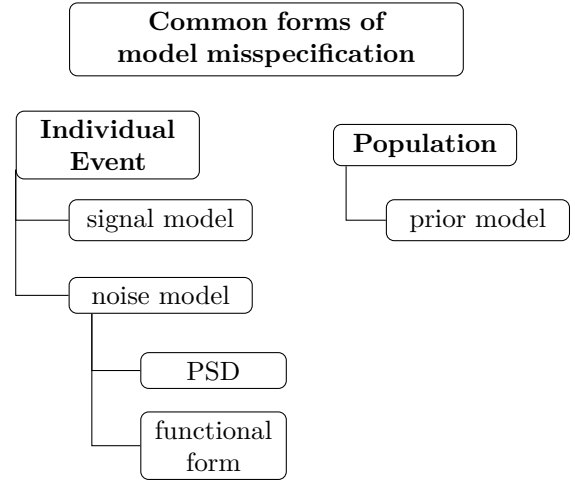
\begin{figure}
\centering
\begin{tikzpicture}
[
  level 1/.style={sibling distance=12em, level distance=4em},
  edge from parent/.style={white}, 
  ]
\node[root] {\textbf{Common forms of model misspecification}}
  child {node[level-2] (c1) {\textbf{Individual Event}}}
  child {node[level-2] (c2) {\textbf{Population}}};
\begin{scope}[every node/.style={level-2}]
\node [below of = c1, xshift=20pt] (c11) {signal model};
\node [below of = c11] (c12) {noise model};
\end{scope}
\foreach \value in {1,2}
  \draw[-] (c1.195) |- (c1\value.west);
\begin{scope}[every node/.style={level-3}]
\node [below of = c12, xshift=20pt] (c121) {PSD};
\node [below of = c121] (c122) {functional form};
\end{scope}
\foreach \value in {1,2}
  \draw[-] (c12.195) |- (c12\value.west);
\begin{scope}[every node/.style={level-2}]
\node [below of = c2, xshift=20pt] (c21) {prior model};
\draw[-] (c2.195) |- (c21.west);
\end{scope}
\end{tikzpicture}
\caption{
Forms of misspecification that we explore in this Article. Individual events can be misspecified if the model for the noise or the signal is not an adequate description of reality. The population of events may also be misspecified. This manifests itself as prior misspecification, which can impact both individual analyses (where the prior may be restricted to a limited portion of the true extent of the posterior) and population analyses (where the goal is to uncover the true distribution of the population).
}
\label{fig:misspecification}
\end{figure}

Models for the data are built on assumptions about the nature of both the noise and signal being measured.
The data model is described by the likelihood function
\begin{align}
    {\cal L}(d | \theta) ,
\end{align}
where $d$ is the data and $\theta$ is a set of parameters describing the noise and/or signal.\footnote{Throughout, we follow the notation from \cite{intro}.}
The likelihood function is a normalised probability density function \textit{for the data}, not for the parameters $\theta$ \citep[see][for more details]{intro}:
\begin{align}
    \int d(d) \, {\cal L}(d| \theta) = & 1 , \\
    \int d\theta \, {\cal L}(d| \theta) \neq & 1 .
\end{align}
It is useful to define a marginal likelihood, which is also known as the Bayesian evidence:
\begin{align}
    {\cal L}(d) = \int d\theta 
    {\cal L}(d|\theta) \pi(\theta) .
\end{align}
Here, $\pi(\theta)$ is the prior distribution for the parameters $\theta$.
The marginal likelihood is a model for the data, averaged over realisations of $\theta$.

A well-known example of a model for gravitational-wave data is the Whittle likelihood model for Gaussian time-series noise, shown here for a single frequency bin:
\begin{align}\label{eq:gaussian_noise}
    {\cal L}(\tilde{d}|\theta) = 
    \frac{1}{2\pi\sigma^2} 
    e^{-|\tilde{d}|^2 / 2\sigma^2} .
\end{align}
Here, $\tilde{d}$ represents the frequency-domain gravitational-wave strain while $\sigma^2$ is related to the noise power spectral density (PSD) $P$ and the frequency bin width $\Delta f$:
\begin{align}
    \sigma^2 = \frac{P}{4\Delta f} .
\end{align}
Equation~\ref{eq:gaussian_noise} is an example of a \textit{parameter-free} model of the data.
If, for example, a gravitational-wave signal from a compact binary coalescence is present, then the likelihood depends on the $\gtrsim15$ parameters associated with a compact binary coalescence (component masses, spins, etc.), 
\begin{align}\label{eq:GW_liklihood}
    {\cal L}(\tilde{d}|\theta) = \prod_k 
    \frac{1}{2\pi\sigma_k^2} 
    e^{-\left|\tilde{d}_k-\tilde{h}_k(\theta)\right|^2 / 2\sigma_k^2} .
\end{align}
Here, $\tilde{h}_k(\theta)$ is a model for the frequency-domain strain from a gravitational-wave signal given binary parameters $\theta$ in frequency bin $k$.
Sometimes, $h_k(\theta)$, which can be defined in either the time domain or the frequency domain, is referred to as ``the waveform model.''

Examining Eq.~\ref{eq:GW_liklihood}, we can see various ways in which the likelihood can be misspecified, which we represent diagrammatically in Fig.~\ref{fig:misspecification}.
First, the waveform $h(\theta)$ may be misspecified, which can lead to well-documented systematic error \citep{Ohme2013,Wade2014,Ashton2020,Gamba2021,Huang2021}.
This is an example of a misspecified signal model.
Second, the noise model can be misspecified.
There are typically two ways that this can happen.
One possibility is that the functional form of the likelihood is correct, but the noise PSD is misspecified.
A number of papers have proposed different methods for estimating the noise PSD in order to minimise this form of misspecification, e.g.,  \cite{Littenberg2015,Cornish2015,Chatziioannou2019}.

The other possibility is that the functional form of the likelihood is itself misspecified.
This may occur because of non-Gaussian noise artefacts \citep{Rover2010} or uncertainty in the noise PSD \citep{student-t,Biscoveanu2020,Banagiri2019}, both of which yield broader tails than the Whittle distribution.
Likewise, marginalising over calibration uncertainty broadens the likelihood function \citep{Sun2020,calibration,Vitale2021}.\footnote{Technically, calibration error is a form of \textit{signal} misspecification in which the gravitational waveform $\lambda(f) \tilde{h}(f)$ includes a calibration correction $\lambda(f)$.
However, marginalising over calibration uncertainty changes the functional form of the likelihood like the other examples in this list.}
Covariance between neighbouring frequency bins induced from finite measurements of continuous noise processes can also lead to misspecification if not correctly accounted for \citep{windows,Isi2021}.
In the subsequent Subsections, we describe tests for these different forms of misspecification.

\subsection{Testing for a misspecified signal model}\label{sec:waveform}
In order to test for a misspecified waveform, it is sometimes useful to look at the whitened\footnote{Whitening is the process by which frequency-domain data $\tilde{d}(f)$ are normalised by the frequency-dependent noise $\tilde{d}(f)\rightarrow\tilde{d}(f)/\sigma(f)$.} residuals of the data in frequency
\begin{align}
    \tilde{r}(f|\theta) \equiv \frac{\tilde{d}(f)-\tilde{h}(f|\theta)}{\sigma(f)} , 
\end{align}
and time
\begin{align}
    r(t|\theta) = {\cal F}^{-1}\left[\tilde{r}(f|\theta)\right] ,
\end{align}
where ${\cal F}^{-1}$ is the discrete inverse Fourier transform.
Residuals can be useful to test for waveform misspecification because the differences between waveform models are clearly seen in the time and frequency domain.\footnote{To see an example of residual analysis from optical astronomy, we direct the reader to the two-dimensional light intensity profiles in \citet{Weinzirl:2008:spectral-decomposition}.}
Additionally, if there is a terrestrial noise artefact (glitch) present in the data, it is likely to appear clearly in the residuals.
For example, in \cite{GW150914}, the best-fit, time-domain residuals for GW150914 were shown to be consistent with Gaussian noise (see Fig.~1 of \cite{GW150914}), helping to show that the data are well explained by a gravitational waveform in Gaussian noise.

\begin{figure}\centering
    \includegraphics[width=0.5\textwidth]{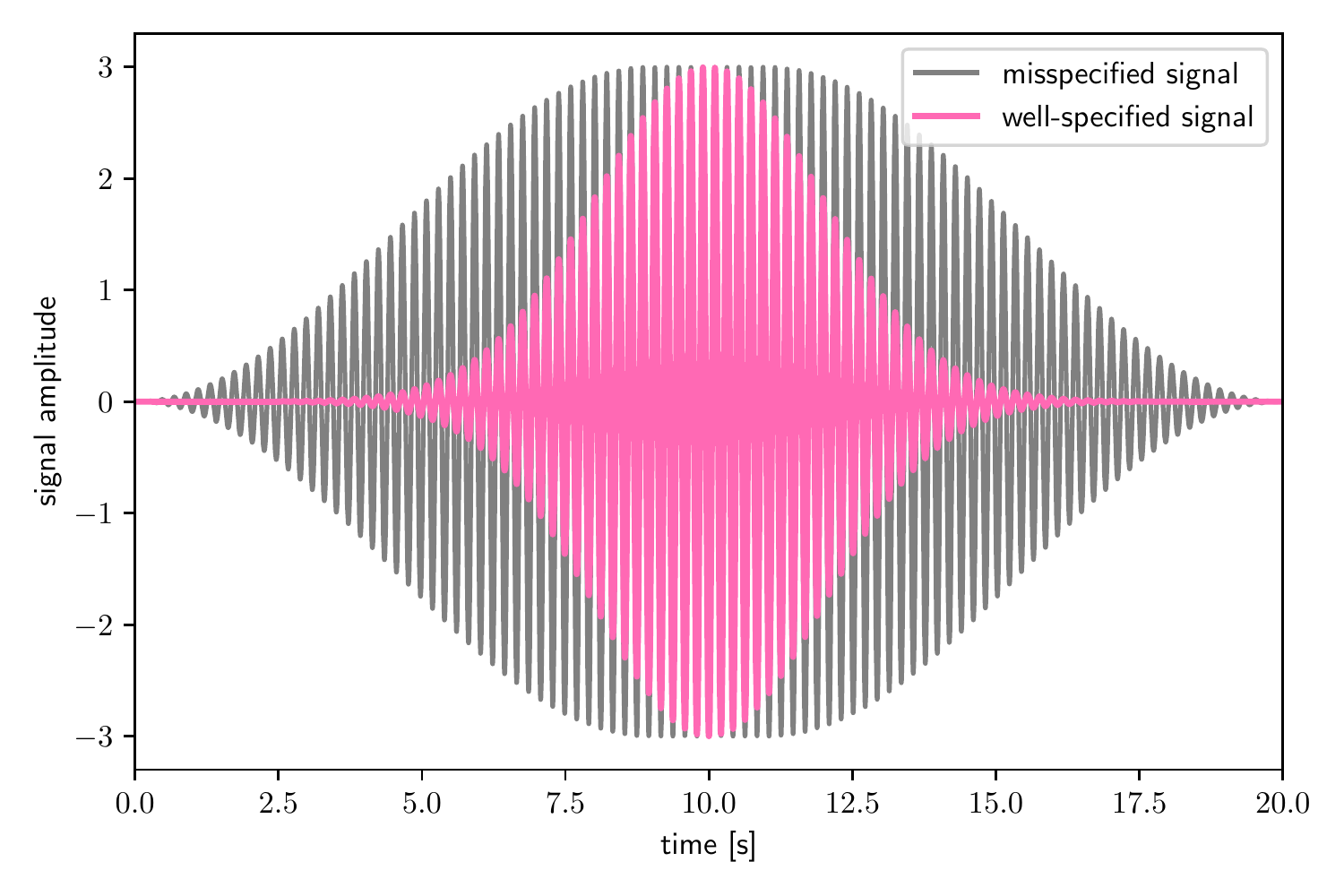}
    \caption{
    The correctly specified waveform (pink) and the misspecified waveform (grey) used in Section \ref{sec:waveform}, plotted in the time domain.
    }
    \label{fig:time_domain_data}
\end{figure}

\textit{Recommended tests:}
\begin{itemize}
    \item Plot the whitened residuals in both time $r(t)$ and frequency $|\tilde{r}(f)|$ and visually inspect for consistency with zero.
    In the frequency domain, the set of $\tilde{r}(f)$ are approximately independent measurements with Gaussian uncertainty equal to unity.
    Include residual curves for many posterior-distribution draws of $\theta$ in order to show theoretical uncertainty. 
    A word of caution: the time-domain residuals are highly covariant in real data, and so it is less straightforward to interpret misspecification in the time domain than in the frequency domain.
    \item As a first step, look for consistency by eye.
    If there are signs of misspecification, one may quantify the inconsistency, e.g., ``the residuals are five standard deviations away from zero at $\unit[500]{Hz}$.''
    While post-hoc analysis is helpful for catching egregious misspecification, in an ideal world, one should define the consistency tests \textit{a priori} for unbiased tests.
    In practice, this is not always possible.
\end{itemize}

\textit{Demonstration:}\footnote{\href{https://colab.research.google.com/drive/1bl5Inibv3qAF0ZUcV-tpCmVFpWxeMu26}{\textcolor{blue}{\textit{Misspecified signal model} notebook}}}
Our signal model is a sine-Gaussian chirplet (i.e., a sine wave multiplied by a Gaussian function).
We create two synthetic sets of data with Gaussian noise.
The \textit{correctly specified} data contains a signal that matches our model.
The second dataset contains an intentionally \textit{misspecified} signal: the same sine wave as before, but multiplied by a Tukey window.
In both datasets, we assume Gaussian noise with a known power spectral density.
In Fig.~\ref{fig:time_domain_data}, we plot these two simulated signals.

It is worth pausing to distinguish this discussion of model misspecification from the similar-but-different topic of model selection.
If we were discussing model selection, we would keep the data fixed and compare it two different signal models.
However, misspecification occurs when the analyst has not conceived of the correct model to test.
Therefore, since we are discussing misspecification, we keep the model fixed and consider two hypothetical datasets: one correctly specified and one misspecified.

\begin{figure*}
    \centering
    \begin{subfigure}[t]{1.0\textwidth}\centering
        \includegraphics[width=0.9\textwidth]{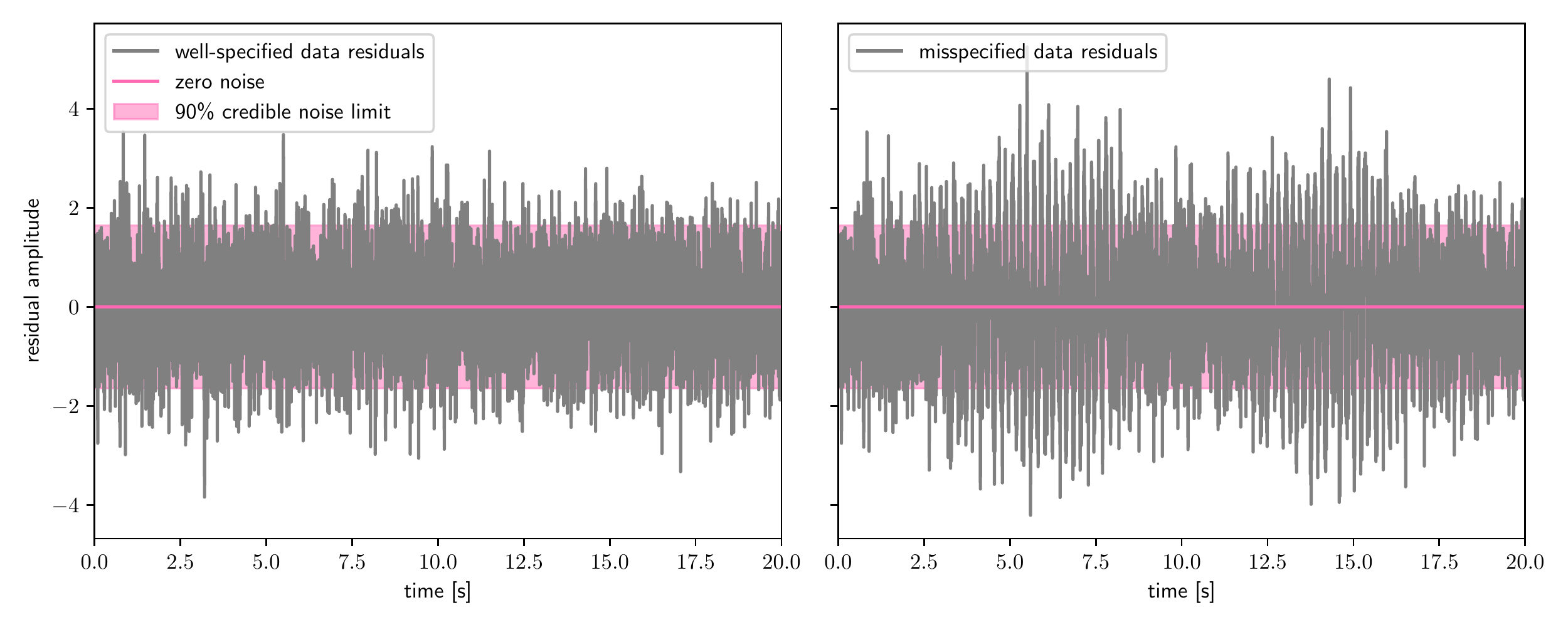}
        \caption{
        Time series of the residuals calculated by subtracting two different waveform models from the simulated data. The plot on the left shows the residuals obtained by subtracting a correctly specified waveform that matches the signal hidden in the data, while those obtained by subtracting a misspecified waveform are on the right.
        \label{fig:time_residuals}}
    \end{subfigure}
    \begin{subfigure}[t]{1.0\textwidth}\centering
        \includegraphics[width=0.9\textwidth]{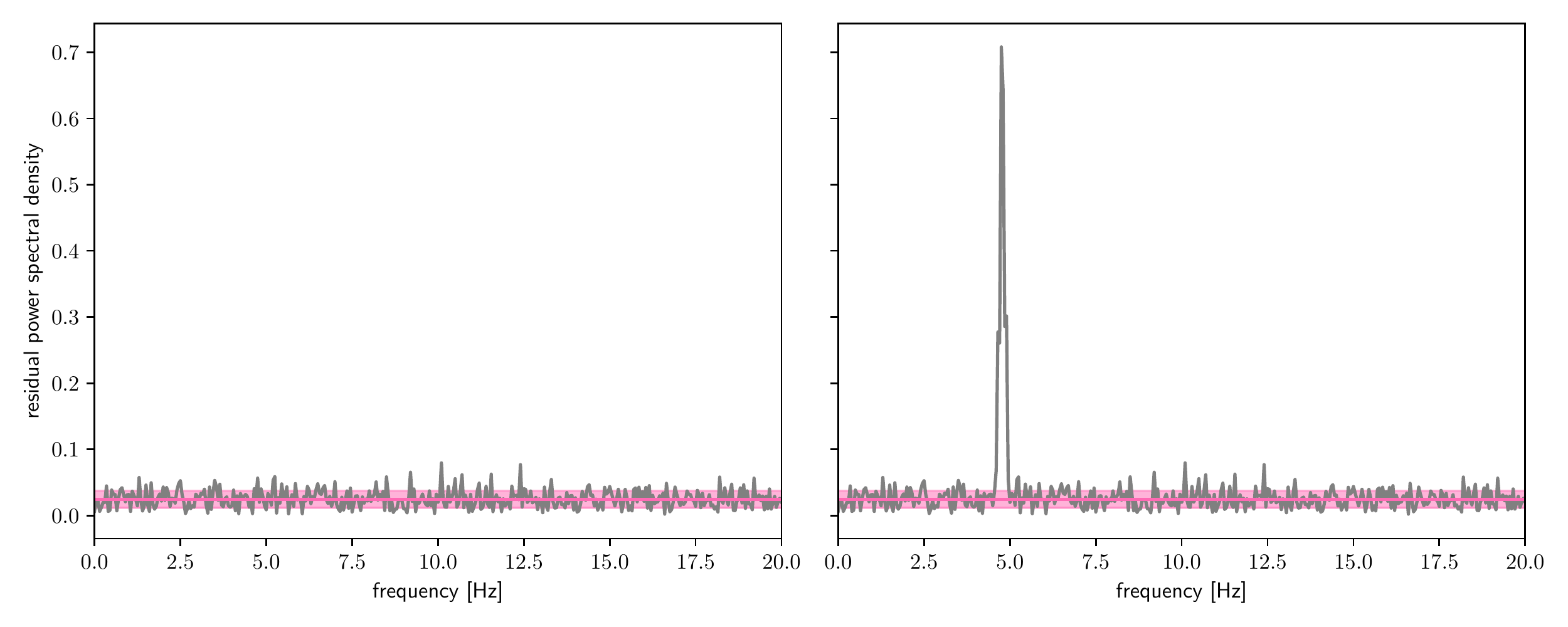}
        \caption{
        Frequency-domain amplitude spectral densities of the residuals.
        It is not totally clear from the time domain data if one of the datasets is poorly specified by the model, but Fourier transforming the residuals reveals a suspicious peak inconsistent with Gaussian noise.
        On both rows, the residuals are plotted in grey while the pink band indicates the range where the model predicts 90\% of the residuals will lie.
        \label{fig:freq_residuals}}
    \end{subfigure}
    \begin{subfigure}[t]{1.0\textwidth}\centering
        \includegraphics[width=0.9\textwidth]{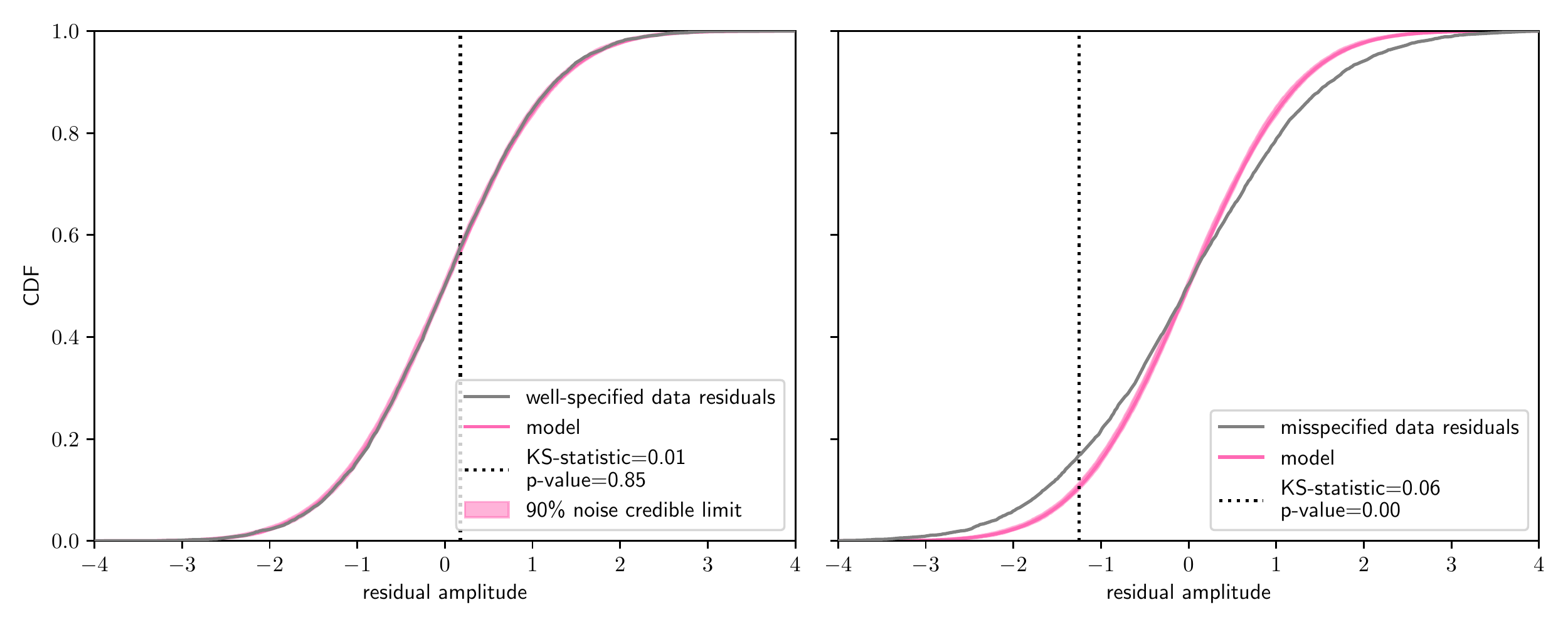}
        \caption{
        Cumulative density functions (CDFs) for the residuals of the time-domain data (grey) and predicted range of the residuals (pink).
        We calculate the KS-statistic for the correctly specified and misspecified distributions. The location at which the KS test finds the maximum vertical distance between the model and the data is indicated with a dotted line. For this specific realisation of Gaussian noise, the KS-statistics are 0.01 and 0.06 for the correctly specified and misspecified data, with respective $p$-values of 0.85 and 0.00 (actually $8.25 \times 10^{-17}$).
        \label{fig:residual_CDFs}}
    \end{subfigure}
    \caption{
    Identifying a misspecified signal model.
    The left-hand column shows tests performed on data containing a signal consistent with the sine-Gaussian pulse model that we test against.
    The right-hand column shows the same tests performed on data containing a different signal.
    }
    \label{fig:residuals}
\end{figure*}

In the right-hand panel of Fig.~\ref{fig:time_residuals}, we display the time-domain residuals obtained when we subtract the sine-Gaussian waveform model from the misspecified data. In the left-hand panel, we show the time-domain residuals obtained from subtracting the same waveform model from the correctly specified data.
If the waveform is correctly specified, the residuals should be consistent with our Gaussian noise model as in the left panel.
Although it is sometimes possible to see misspecification in the time-domain data (for example, when there is a short glitch), the misspecification may be more apparent in the frequency domain as is the case here.

In Fig.~\ref{fig:freq_residuals}, we plot the amplitude spectral density of the residuals. 
Again, the left-hand panel shows the residuals for the correctly specified signal while the right-hand panel shows the residuals for the misspecified signal.
While the correct waveform yields residuals consistent with the noise amplitude spectral density, the misspecified signal produces a peak in the amplitude spectral density that is conspicuously outside of the 90\% range predicted by the model (shown in pink).

In order to assess the overall goodness of fit, it can be useful to plot the cumulative density function (CDF) of the residuals alongside the theoretical CDF predicted by the model.
The CDF may be constructed from residuals in the time domain (to highlight transient phenomena), but more often, misspecification is most apparent using whitened frequency-domain residuals.
An illustration of the CDF test is provided in Fig.~\ref{fig:residual_CDFs}, where we plot the CDF of the time-domain residuals.
While there is only one realisation of the data (grey), we can generate arbitrarily many realisations of the theoretical CDF (pink).
The thickness of the pink CDF shows the variability from $100$ different realisations.\footnote{This is similar to, although not the same as, Gelman's posterior predictive checks \citep[e.g.][]{Gelman, GelmanShalizi:2010}. Gelman's checks involve drawing realisations from the histogram of the posterior probability distribution under the assumption of the model, and checking how probable it is for the model to produce realisations that are consistent with the observed data. This is something that we return to in Section \ref{priors}. Here, we draw realisations from the noise model in order to establish the range over which it can credibly vary.}

To determine if the residuals agree with the model, one can employ any of the many established hypothesis-testing tools available to determine if measured samples are drawn from the theoretical distribution.
For example, the Kolmogorov-Smirnov (KS) statistic,
\begin{align}
    D_n = \max_f\Big|
    \text{CDF}(f|\text{measured}) - 
    \text{CDF}(f|\text{predicted})
    \Big| ,
\end{align}
is the maximum difference between the measured CDF of the data and the predicted CDF given $n$ frequency bins.
(The Anderson-Darling test is also commonly used to determine if measured samples are drawn from a predicted distribution.)
The $D_n$ statistic can be converted into a $p$-value with a look-up table that does not depend on the functional form of the CDF.
The Kolmogorov-Smirnov $p$-values for the correctly specified data and misspecified dataset are provided in the legends of Fig.~\ref{fig:residual_CDFs}.
The small $p$-value in the right panel suggests that our signal model is indeed misspecified.

Sometimes one may wish to inspect the data within a particular time or frequency interval.
In this case, it can be enlightening to draw hundreds of distributions from the model, and count the number of times that the model CDF in this bin is above the CDF of the data.
The fraction of draws above the CDF in this bin constitutes a $p$-value; if the data are correctly specified, the $p$-value will follow a uniform distribution.
Thus, a $p$-value very close to $0$ or $1$ indicates that the data in this bin deviate significantly from the model.
However, care must be taken if more than one $p$-value is calculated in the same set of data---while each $p$-value is uniformly distributed, the set of $p$-values from one dataset are correlated with each other.

Both of the tests described above assume that the predicted model is non-parametric.
That is, the distribution of the residuals does not depend on any parameters.
If the predicted distribution depends on one or more parameter $\theta$, then we must rely on Monte Carlo methods to determine if the distributions agree.
For example, we may calculate
\begin{align}
    D'_n = \max_f\Big|
    \text{CDF}(f|\text{meas.}) - 
    \text{CDF}(f|\widehat\theta,\text{pred.}) \Big| ,
\end{align}
which is a KS-type statistic using the maximum likelihood parameters $\widehat\theta$.
Since we use the data to estimate $\widehat\theta$, $D'_n$ is not distributed according to the Kolmogorov distribution.
(It is easier to get smaller differences between the measured and predicted CDFs when the theoretical CDF varies depending on the parameters.)
However, we can still calculate a $p$-value by generating synthetic data to empirically estimate the distribution of $D'_n$, which amounts to a generalisation of Lilliefors test \citep{Lilliefors1967}.
We demonstrate such a calculation below in Subsection~\ref{parameterised_priors} in the context of prior misspecification; see in particular Fig.~\ref{fig:ks-dist-hist}.

\subsection{Testing for a misspecified noise model}\label{sec:non-whittle}

In order to observe noise misspecification and deviations from the Whittle likelihood, it is again useful to look at the  distributions of residuals.
By comparing the observed residuals to the theoretical likelihood distribution, it is possible to see, for example, if five-sigma deviations are more common than expected.

\textit{Recommended tests:}
\begin{itemize}
    \item Create a histogram representing the probability density function of the data; or, alternatively, plot the CDF of $r,|\tilde{r}|$, the whitened residuals.
    It is sometimes also useful to plot the distribution of the whitened residual power $|\tilde{r}|^2$; see~\cite{student-t,Chatziioannou2019}.
    Include residual curves for many posterior-distribution draws of $\theta$ in order to show theoretical uncertainty.
    \item For fixed models with no parameters, calculate a goodness-of-fit statistic using, for example, the KS test.
    When models have parameters, calculate the goodness of fit for the maximum-likelihood parameters.
    \item Plot the difference in the cumulative density functions (empirical - expected) as a function of the expected CDF as in \cite{student-t, Chatziioannou2019}.
    This is like a probability--probability (``$PP$'') plot, in which the fraction of repeated measurements is plotted against the confidence level at which the known truth value exists, for data model checking.
    \item Bootstrapping methods---in which real data is used as a sampling distribution for synthetic data---can often be helpful for diagnosing noise misspecification. In gravitational-wave astronomy, data residuals can be bootstrapped to create new noise realisations; this method was integral to the first gravitational-wave detections \cite{CannonTimeslides, Cannon2015, AshtonBootstrap}. New noise realisations may also be generated from existing data using methods like time-sliding or time reversal. However, all bootstrap methods ultimately break down due to saturation effects \citep{Was2010}; it is impossible to simulate all possible noise realisations using a finite dataset. In astro-particle physics, using instruments such as Super-Kamiokande, sidereal time scrambling is used to estimate typical fluctuations in signal strength due to noise \citep{ThraneParticle}.
\end{itemize}

\textit{Demonstration:}\footnote{\href{https://colab.research.google.com/drive/1ZdYAoFPrYWfFSlUa8exXGBsGJDOlY4p7}{\textcolor{blue}{\textit{Misspecified noise model} notebook}}}
We demonstrate how to diagnose noise misspecification.
The true noise is Gaussian with a mean $\mu=0$ and standard deviation $\sigma=1$.
The misspecified noise is distributed according to the Student's $t$ distribution with $\nu=5$ degrees of freedom.
These parameters are chosen so that the noise profiles appear, at first glance, to be consistent with each other.

\begin{figure*}
    \centering
    \begin{subfigure}[t]{1.0\textwidth}\centering
        \includegraphics[width=0.9\textwidth]{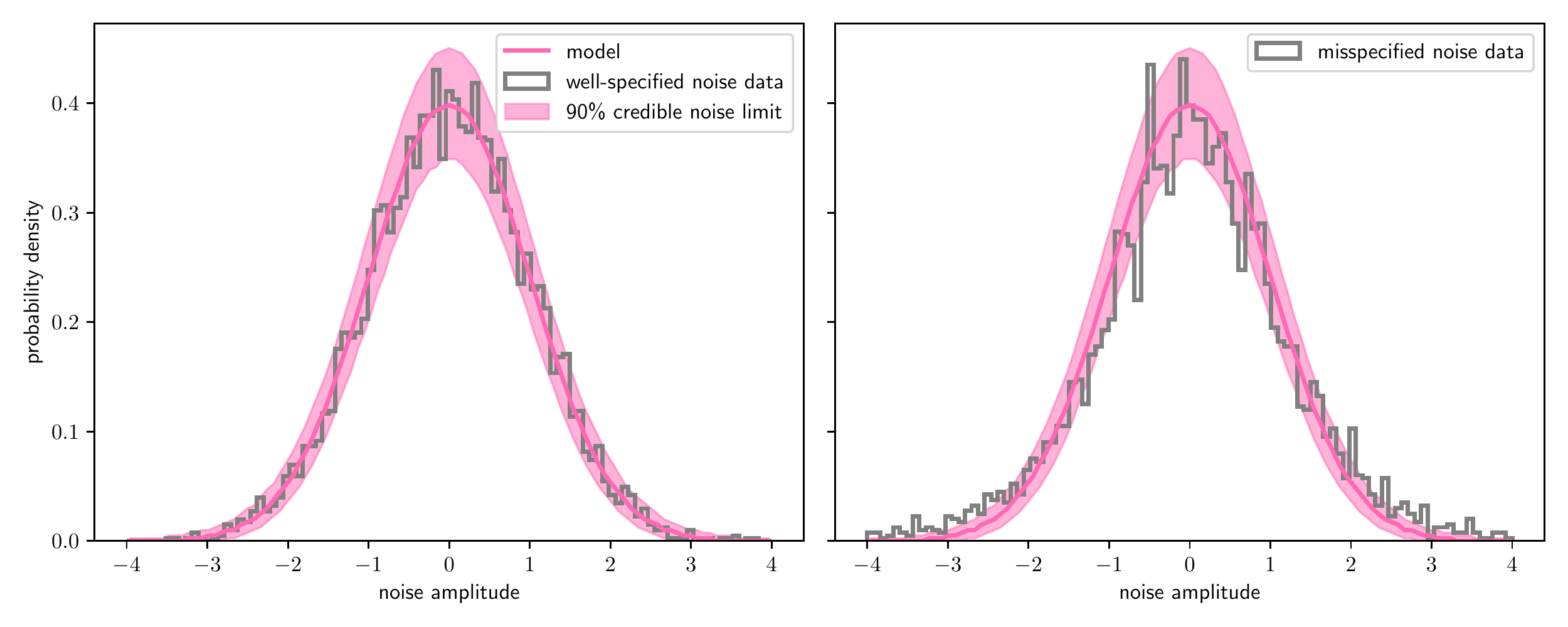}
        \caption{\label{fig:noise_histogram_frequency}Simulated noise distributed as a Gaussian in the frequency domain. The correctly specified Gaussian distribution (left) and the similar-but-misspecified Student's $t$ distribution (right). The predicted (90\% credible) range predicted by the model is shown by the pink band.
        }
    \end{subfigure}
        \begin{subfigure}[t]{1.0\textwidth}\centering
        \includegraphics[width=0.9\textwidth]{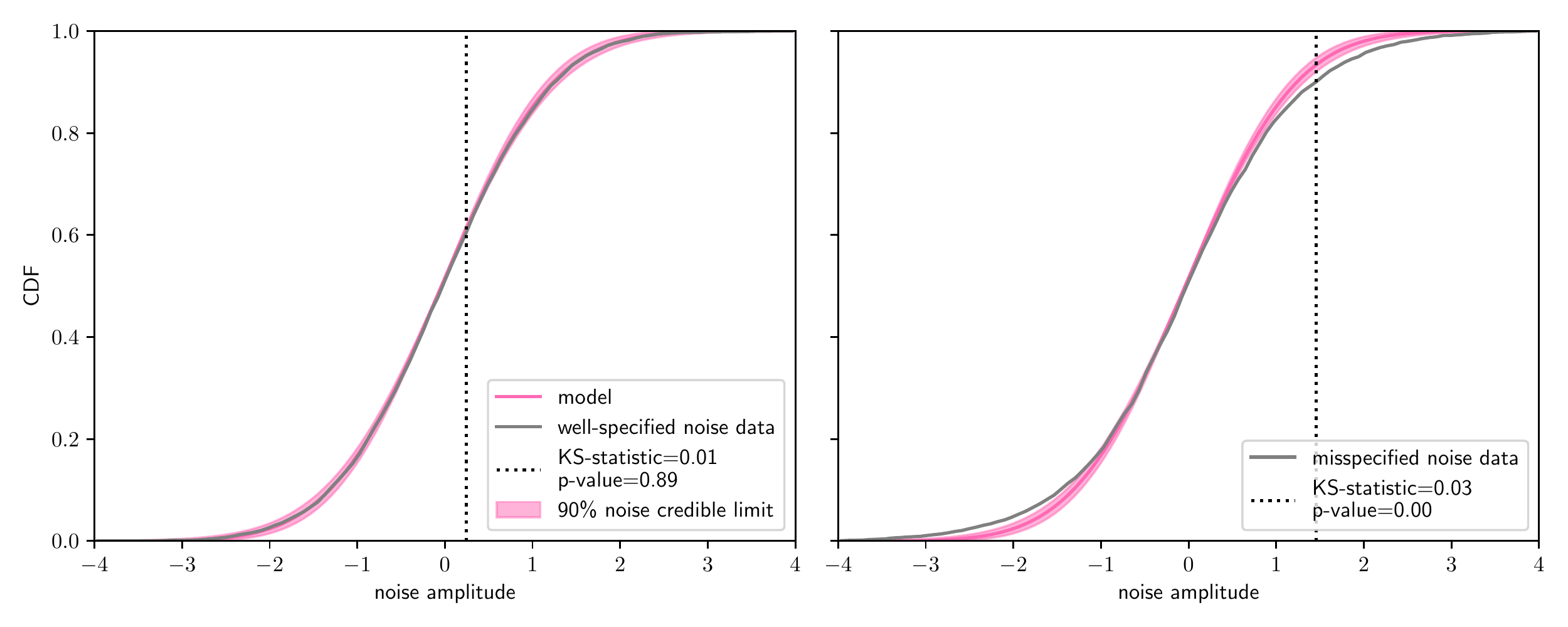}
        \caption{\label{fig:noise_histograms:cdf}
        Cumulative density functions (CDFs) for frequency-domain residuals (grey) and the range predicted by the model (pink).
        For these specific noise realisations, the KS-statistics are 0.01 and 0.03 for the correctly specified and misspecified models, with $p$-values of $0.89$ and $3.03 \times 10^{-6}$ respectively. 
        As in Figure \ref{fig:residual_CDFs}, the location of the maximal KS distance is noted by a dotted line.
        }
    \end{subfigure}
    \begin{subfigure}[t]{1.0\textwidth}\centering
        \includegraphics[width=0.9\textwidth]{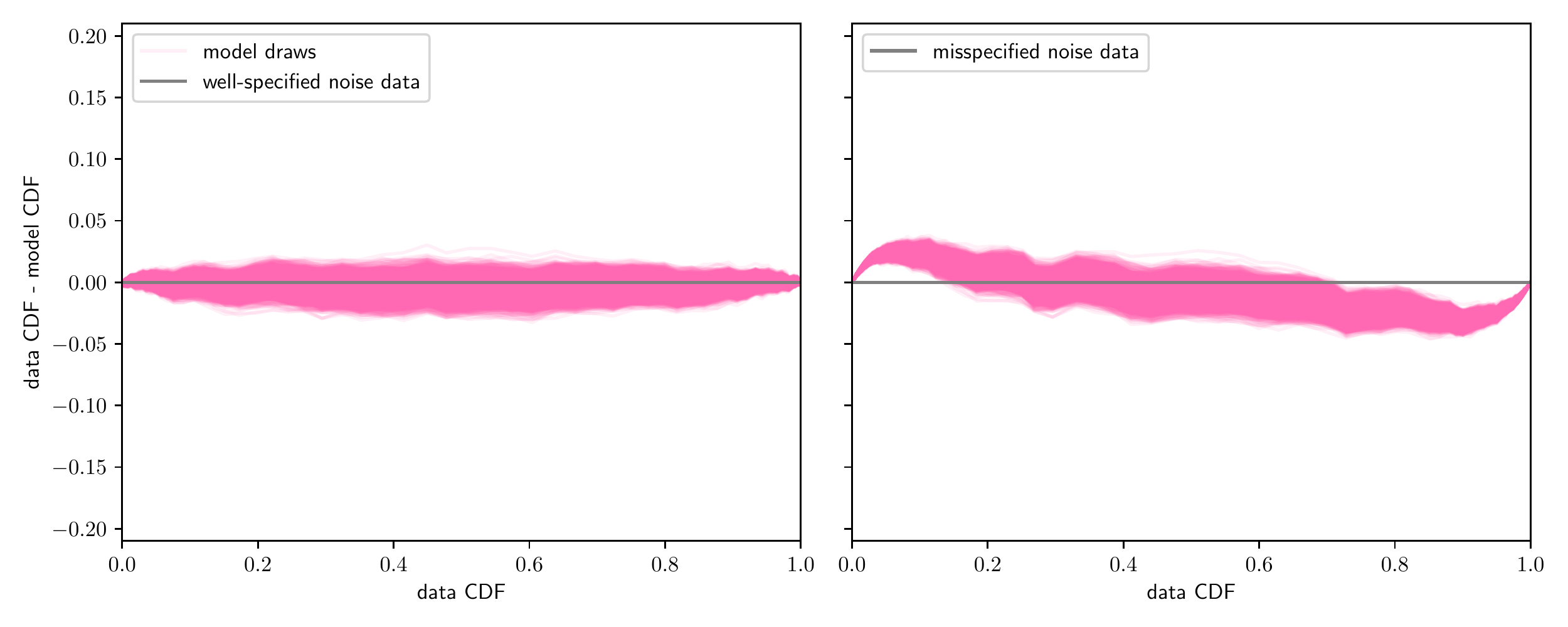}
    \caption{\label{fig:noise_histograms:pp}
    Like the previous row, but the \textit{difference} in data and model CDFs as a function of the data CDF.
    }
    \end{subfigure}
    \caption{
    Identifying a misspecified noise model.
    The left-hand column shows tests performed with a correctly specified Gaussian noise model while the right-hand column shows the same tests with the same Gaussian model, but performed against a misspecified Student's-$t$ distribution.
    }
    \label{fig:noise_histograms}
\end{figure*}

We display histograms of each noise dataset in Fig.~\ref{fig:noise_histogram_frequency}.
The region that the model predicts 90\% of the data to lie within is also shown in these plots.
The histograms both appear to largely lie within this 90\% credible region, with the Student's $t$-distributed data only visibly deviating in the tails.

Next, we Fourier transform our datasets and compare the 90\% range predicted by the noise model against histograms of the data in the frequency domain (Fig.~\ref{fig:noise_histogram_frequency}).
In the frequency domain, the misspecified data more clearly strays outside of the range predicted by the model.
We then create a CDF of the frequency-domain data, comparing again to the 90\% range predicted by the model (Fig.~\ref{fig:noise_histograms:cdf}).
We find a reasonable KS-test $p$-value for the correctly specified data and an extreme $p$-value for the misspecified data, as expected.
Finally, we compare the data to the model by plotting the data CDF against the difference between the data CDF and the model CDF (Fig.~\ref{fig:noise_histograms:pp}).
In this final test, it is clear that the data is not well-represented by the model.

\section{Prior misspecification}\label{priors}
In the previous Section, we discussed various ways in which the likelihood can be misspecified and how we can detect this misspecification.
Now we turn our attention to the misspecification of the \textit{prior}, $\pi(\theta)$, a distribution that describes our prior knowledge of the parameters $\theta$.

\subsection{Priors with no hyper-parameters}
In some cases, we can be very confident in our priors.
For example, since there is no preferred direction in the Universe, the best prior for inclination angle $\iota$ (the angle between the orbital angular momentum and the line of sight) is uniform in $\cos\iota$.
In other situations, we are not confident in the form of the prior distribution.
In these cases, it can be useful to formalise this theoretical uncertainty using a conditional prior
\begin{align}
    \pi(\theta | \Lambda) .
\end{align}
Here, $\Lambda$ is a ``hyper-parameter'' we may vary to alter the shape of the prior for $\theta$.
In this Subsection, we focus on priors with no hyper-parameters while we cover parameterised priors in the next Subsection.

\textit{Recommended tests:}
\begin{itemize}
    \item Make a CDF plot comparing the reconstructed distribution of $\theta$ to the expected distribution of $\theta$. 
    In order to obtain a reconstructed distribution, obtain posterior samples for $N$ different events, each \textit{weighted with the population model to be tested}.
    Draw one posterior sample from each of the $N$ events to make a realisation of the reconstructed distribution.
    Do this many times to make many realisations.
    Plot CDFs of the many realisations alongside the population model being tested.
    If the data agrees with the model, the CDFs should overlap.
    If the two CDFs do not overlap, the model is a poor description of the data.
    It is worth noting that a model can pass this test while still being quite badly misspecified. \citet{GWTC-2-pop} found evidence for negatively aligned spins in a population of compact-object binaries, but this was later shown to be a model-dependent feature \citep{Roulet:2021:spin, BuildingBetterModels}. This test may be particularly unreliable if there is a sharp feature in the data that is not captured by the prior.
    \item If a more quantitative test is desired, one may calculate the KS statistic for each CDF. 
    If the best-fitting reconstruction has a $p$-value below a certain threshold, for example $p$-value$\leq0.00005$ \citep{redefine-stats}, then the model is not a good fit to the data. 
    \item One may also identify problem areas of model under- or over-production by quantifying the discrepancies between the data-draw CDFs and the model over the parameter space. This facilitates statements like ``99\% of the time, the model produces too many binary black hole events with $m_1>80 M_\odot$''.
\end{itemize}

\begin{figure*}
    \centering
    \begin{subfigure}[t]{1.0\textwidth}\centering
    \includegraphics[width=0.8\textwidth]{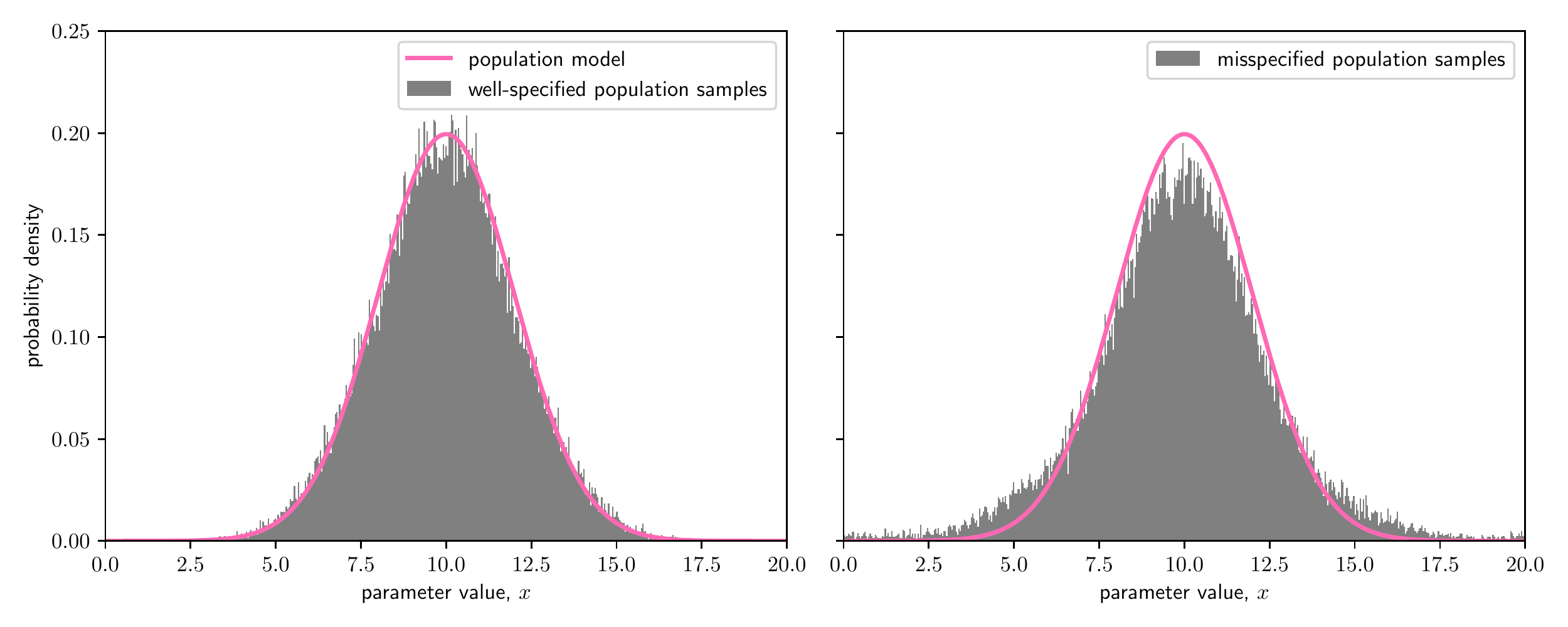}
    \caption{The distribution of the parameter $x$. Simulated posterior samples are grey while the prior distribution used for the construction of the posterior is shown in pink. On the left, the (Gaussian) prior is correctly specified, and on the right, the  Laplacian prior is misspecified.}
    \label{fig:population_models}
    \end{subfigure}
        \begin{subfigure}[t]{1.0\textwidth}\centering
    \includegraphics[width=0.8\textwidth]{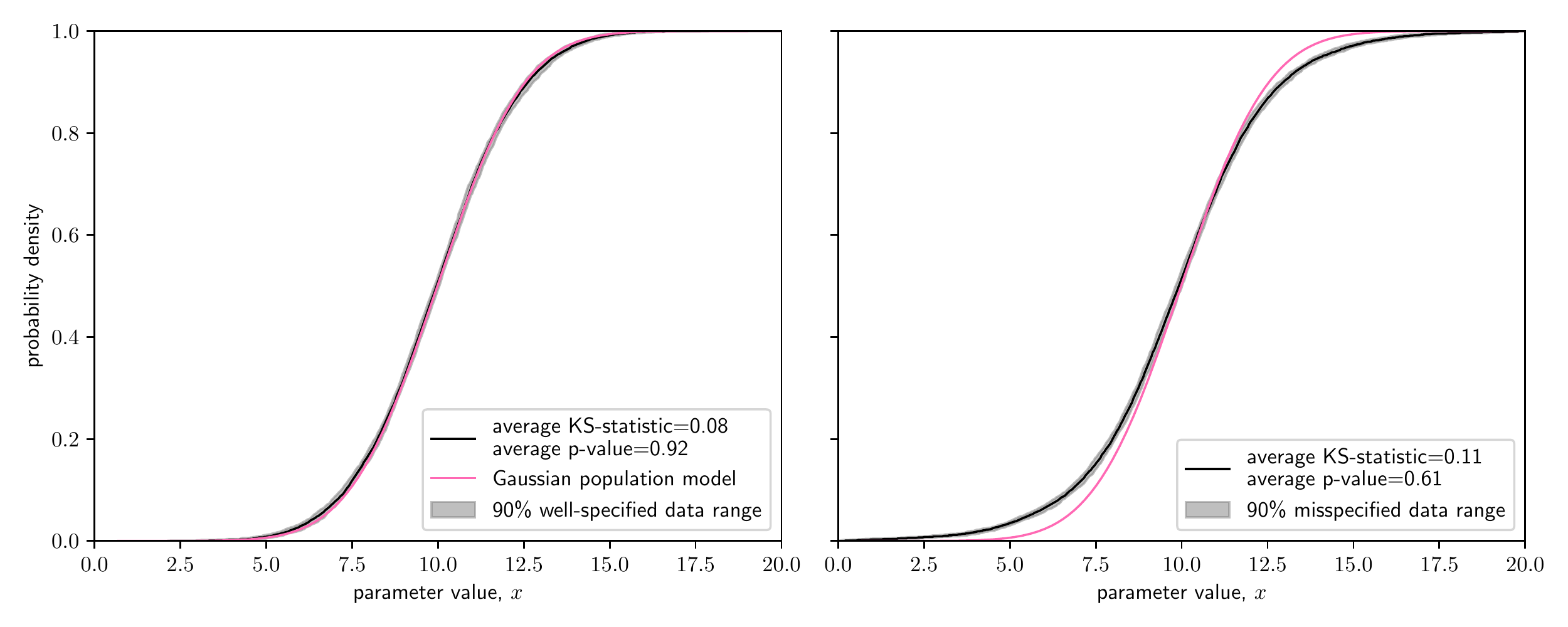}
    \caption{Cumulative distribution of the parameter $x$.
    Pink (observed) shows the 90\% interval for CDFs using draws from the posterior samples.
    Grey (predicted) is $90\%$ predicted range of the CDF of the prior distribution used to generate the posterior samples.
    The misspecified data (right-hand panel) visibly disagrees with the model. The $p$-value for the average of the draws from the CDF is relatively low and the corresponding KS-statistic of the average CDF is relatively high; together with the visible deviation of the model from the data, these values indicate that the model does not well-specify this data.
    }
    \label{fig:population_models_cdf}
    \end{subfigure}
    \begin{subfigure}[t]{1.0\textwidth}\centering
    \includegraphics[width=0.83\textwidth]{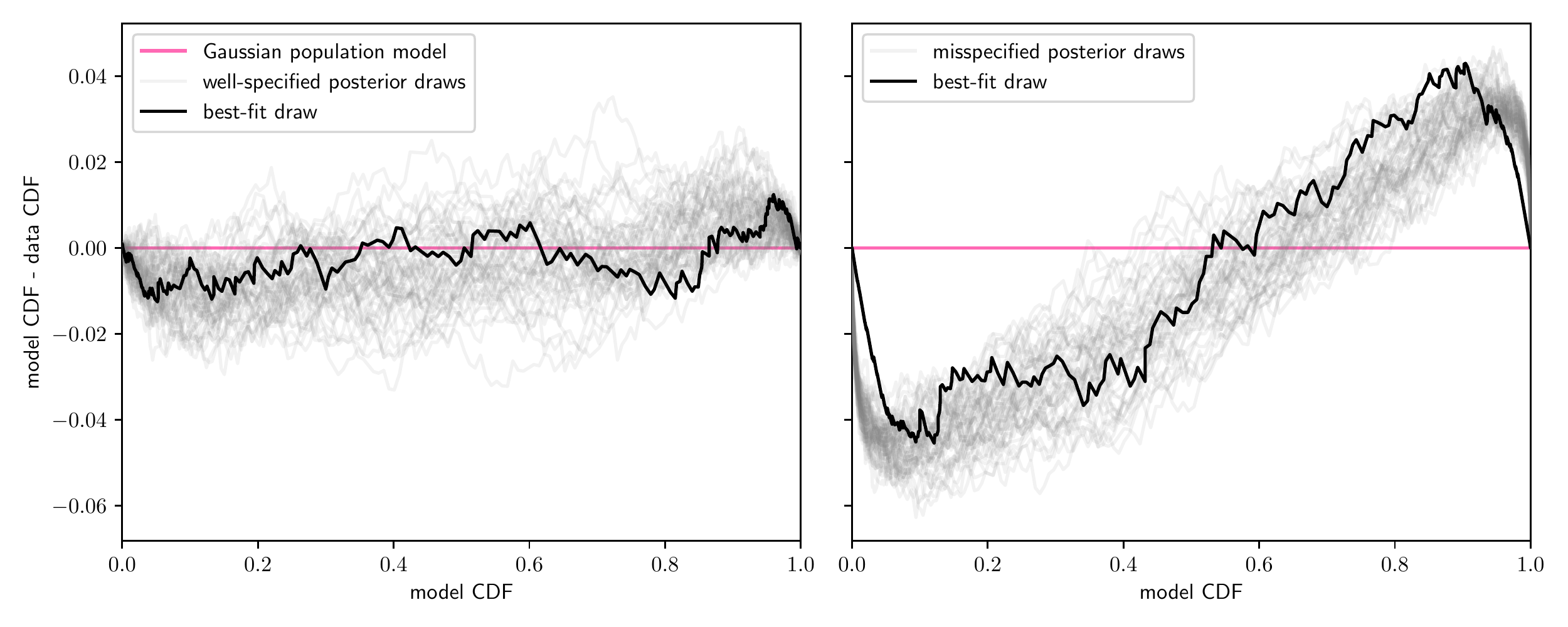}
    \caption{
    The difference between the observed and predicted CDFs as a function of the predicted CDF.
    The difference is inconsistent with zero when the predicted CDF is $\approx 0.1$ and $\approx0.9$}
    \label{fig:population_pp}
    \end{subfigure}
    \caption{Plots illustrating how model misspecification may manifest for an unparameterised population model, with a correctly specified dataset shown in the left-hand plots, and the misspecified case in the right-hand plots.}
\end{figure*}

\textit{Demonstration.}\footnote{\href{https://colab.research.google.com/drive/1rDfDCWKXb7wjHEn3Ktt4SGFvMlUpUCzi}{\textcolor{blue}{\textit{Misspecifided prior: unparameterised case} notebook}}} 
We simulate data $d$ consisting of some physical parameter $x$ and noise $n$:
\begin{align}
    d = x + n .
\end{align}
The noise is Gaussian distributed with zero-mean and unit variance.
The values of $x$ are drawn from the true prior distribution: a Gaussian distribution with mean $\mu=10$ and width $\sigma=2$.
However, in order to demonstrate prior misspecification, we employ as our prior a Laplacian with the same mean and variance as the true distribution.
The true prior distribution and the misspecified prior distribution are compared in Fig.~\ref{fig:population_models}.

We create a simulated dataset of $N=1000$ events.
In each case, we construct Gaussian likelihood curves of mean $\mu_i = d_i$. 
The posterior for each value of $x_i$ is proportional to the likelihood multiplied by the prior.
We draw $50$ posterior samples from each of these simulated posteriors in order to create $50$ CDF curves; see Fig.~\ref{fig:population_models_cdf}.
We make two versions of this plot: one with the correctly specified prior and one with the misspecified prior.

It is also sometimes useful to plot the data CDF minus the model CDF.
We include an example of this plot in Fig.~\ref{fig:population_pp}.
In this case, we see that the misspecified prior yields disagreements in the CDF in the distribution tails; the difference between pink model and grey data is, in general, inconsistent with zero when the model CDF is $\approx0.1$ and  $\approx0.9$.

\subsection{Hyper-parameterised priors}\label{parameterised_priors}

\begin{figure*}
    \centering
    \begin{subfigure}[t]{1.0\textwidth}\centering
        \includegraphics[width=0.9\textwidth]{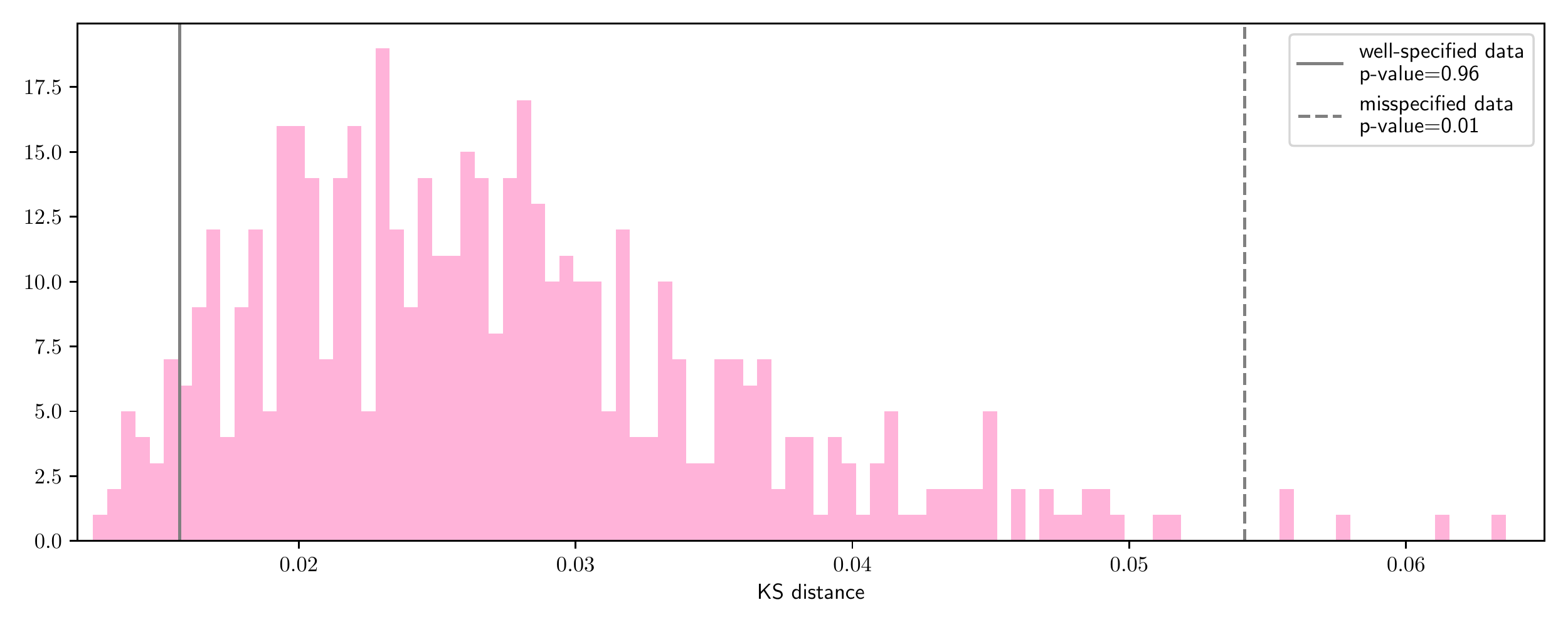}
        \caption{\label{fig:ks-dist-hist}Distribution of KS distances drawn from realisations of the data-conditioned model (pink). We use these to compute a KS-like $p$-value by calculating the fraction of the distribution above the KS distance calculated for the data (grey).
        }
    \end{subfigure}
        \begin{subfigure}[t]{1.0\textwidth}\centering
        \includegraphics[width=0.9\textwidth]{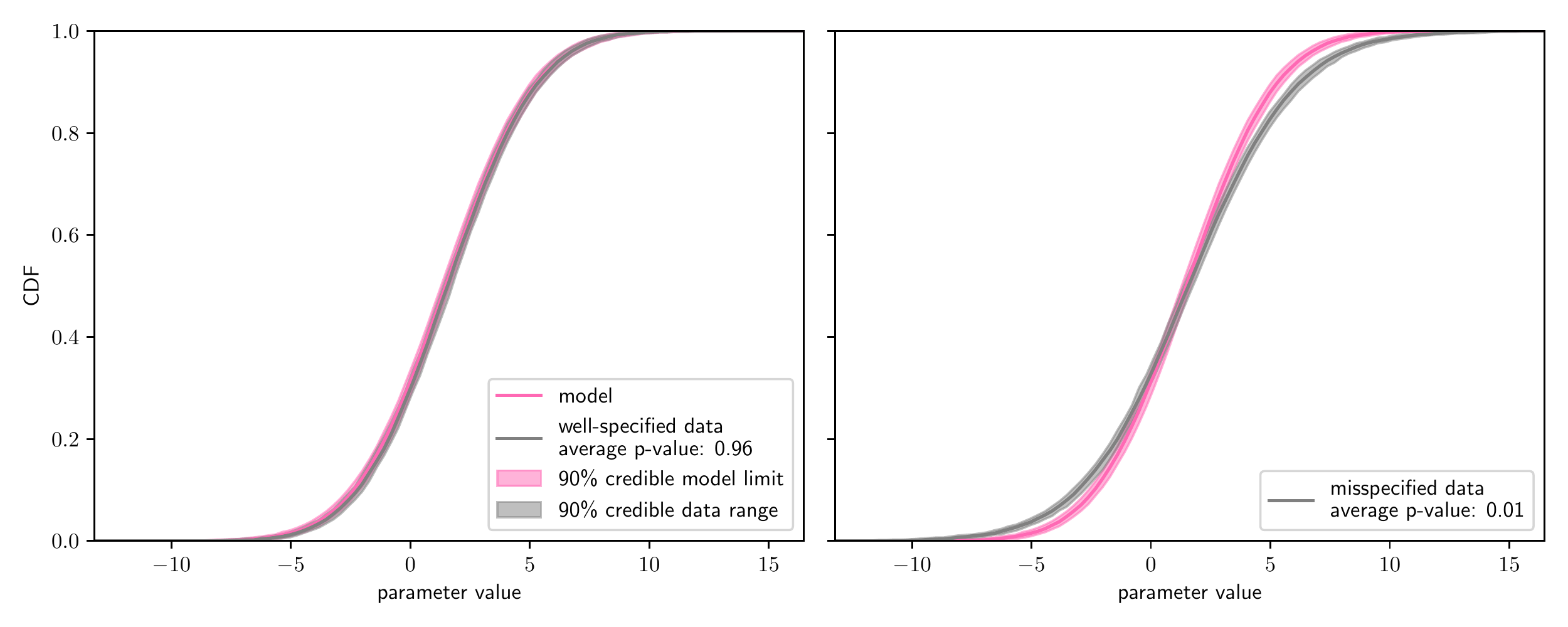}
        \caption{\label{fig:parameterised_pop:cdf}Parameterised model (pink) and the data (grey); the width of both translucent bands represents the 90\% credible interval around that quantity, with the uncertainty coming from the unknown (but estimated) hyper-parameter $\mu$. The misspecified data on the right-hand side clearly strays outside of the 90\% credible bands of the model, even though the model was conditioned on the same data, because the model does not describe the data well.
        }
    \end{subfigure}
    \begin{subfigure}[t]{1.0\textwidth}\centering
        \includegraphics[width=0.9\textwidth]{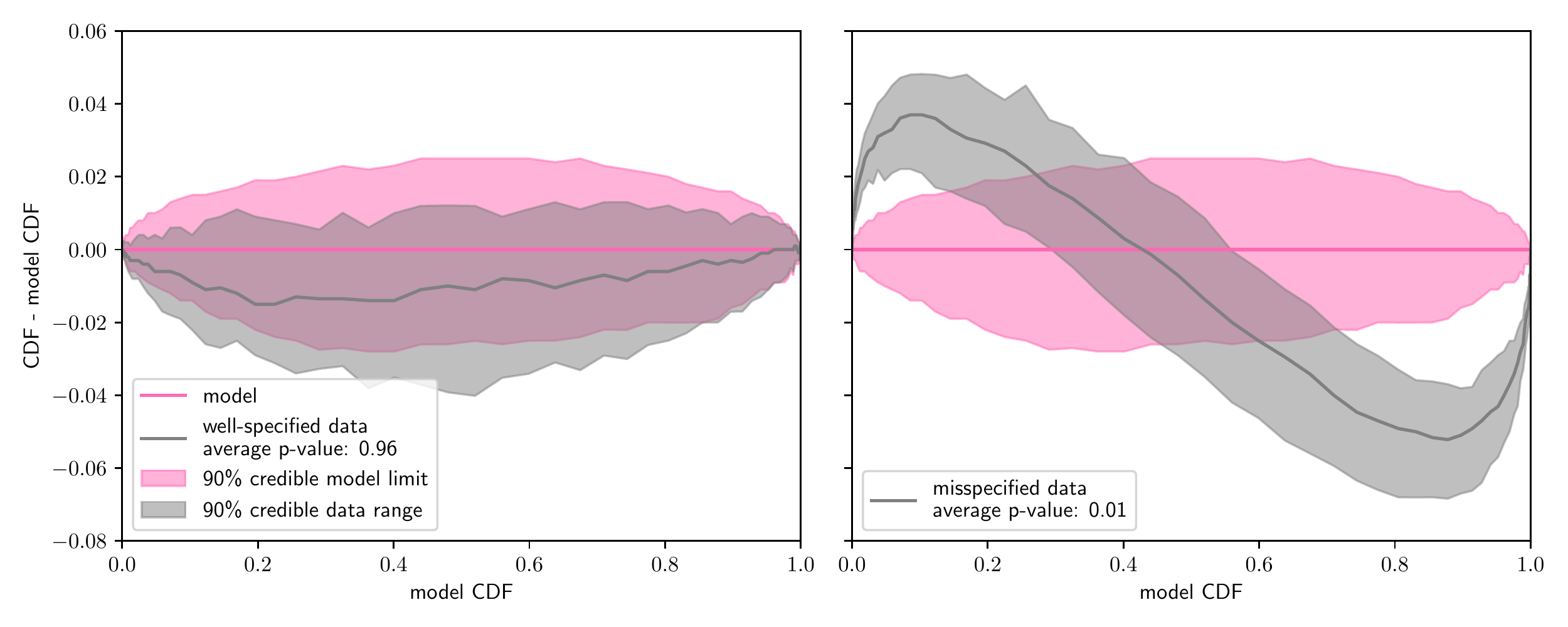}
    \caption{\label{fig:parameterised_pop:pp}
    The difference in data and model CDFs as a function of the model CDF. The median of the correctly specified data (left) stays within the band of 90\%  model uncertainty, but the misspecified data clearly deviates.
    }
    \end{subfigure}
    \caption{
    Model misspecification for a parameterised population model, with the correctly specified case demonstrated in the left-hand side of the lower two plots, and the misspecified case demonstrated on the right.
    }
    \label{fig:parameterised_population}
\end{figure*}

We are particularly concerned with population models (sometimes called ``hierarchical models''), where the prior for the parameters $\theta$ is conditional on a set of hyper-parameters $\Lambda$, describing the shape of the prior distribution:
\begin{align}
    \pi(\theta | \Lambda) .
\end{align}
While data models are built on assumptions about the nature of noise, population models are built on assumptions about astrophysics.
For example, if we assume that the distribution of primary black-hole mass $m_1$ follows a power-law distribution with spectral index $\alpha$,
\begin{align}
    \pi(m_1 | \alpha) \propto m_1^\alpha ,
\end{align}
then $\alpha\in\Lambda$ is a hyper-parameter for the distribution of $m_1 \in \theta$.

Hyper-parameters are frequently a convenient way of describing systematic theoretical uncertainty.
For example, consider a parameter $x$ drawn from a Gaussian distribution ${\cal N}(x|\mu,\sigma)$ with mean $\mu$ and width $\sigma$.
If we do not know the precise value of $\mu$ or $\sigma$, our misspecification tests should take this uncertainty into account.
We repeat the tests from the previous Subsection for the case of a hyper-parameterised prior.

Here is a summary of the differences that arise from the addition of a hyper-parameter.
One must generate random draws of the hyper-parameter $\Lambda$.
Then generate draws of the model CDF for each random draw of $\Lambda$, and weight posterior samples using the population predictive distribution---the conditional prior, marginalised over uncertainty in the hyper-parameters:
    \begin{align}
        \text{ppd}(\theta) = \int d\Lambda\,
        \pi(\theta | \Lambda) 
        p(\Lambda | d) .
    \end{align}
Here $p(\Lambda | d)$ is the posterior for the hyper-parameters.
Take care not to double-count; the posterior for $\Lambda$ used to reweight event $k$ should not be informed by event $k$; see, e.g., \cite{Essick:2021:outliers}.

\textit{Demonstration:}\footnote{\href{https://colab.research.google.com/drive/1Pw158cBHME_2iqmc6xdCspevGyB1l61d}{\textcolor{blue}{\textit{Misspecified prior: parameterised case} notebook}}} 
We assume a Gaussian prior for parameter $x$ with uncertain $\mu$ and width $\sigma$.
We simulate $N=1000$ true events, $x_\mathrm{true}$, from two populations (priors): one Gaussian-distributed and one Laplacian-distributed, both with the same means $\mu_\mathrm{G}$, and with widths $\sigma_\mathrm{G}$ and $\sigma_\mathrm{L}$, respectively.
For each population, we calculate the maximum-likelihood detected value of each event, $x_\mathrm{meas}$ by offsetting it by a random number drawn from a Gaussian of the same width as the likelihood distribution, $\sigma_\mathrm{meas}$.
The mean average of these maximum-likelihood values is the maximum-likelihood estimate for the population prior mean, $\mu_\mathrm{E}$, and the measured population prior width is $\sigma_\mathrm{E}=\sigma_\mathrm{meas} / \sqrt{N}$.
The estimated mean of the population posterior is described by
\begin{equation}
    \mu_\mathrm{P} = \frac{\sigma_\mathrm{E}^{-2} \mu_\mathrm{E} + \sigma^{-2} \mu}{(\sigma_\mathrm{E}^{-2} + \sigma^{-2})} = \frac{\mu_\mathrm{E}}{\sigma_\mathrm{E}^2 (\sigma_\mathrm{E}^{-2} + \sigma^{-2})},
\end{equation}
since $\mu = 0$. The estimated width is
\begin{equation}
    \sigma_\mathrm{P} = (\sigma_\mathrm{E}^{-2} + \sigma^{-2})^{-\frac{1}{2}}.
\end{equation}
The width of the posterior \textit{predictive} distribution is 
\begin{equation}
    \sigma_\mathrm{pp} = \sqrt{\sigma_\mathrm{P}^{2} + \sigma^{2}},
\end{equation}
the width of a cross product of the uncertainty distribution of width $\sigma$ and the posterior width $\sigma_\mathrm{P}$. The population-weighted width for each individual event's posterior is
\begin{equation}
    \sigma_\mathrm{x} = (\sigma_\mathrm{E}^{-2} + \sigma_\mathrm{pp}^{-2})^{-\frac{1}{2}}.
\end{equation}
We draw $M=100$ samples from each of these posteriors of mean $x_\mathrm{meas}$ and width $\sigma_\mathrm{x}$. 
In practice, we do this by drawing $N$ samples $M$ times from the population posterior.
The CDFs of these draws can then be compared to the model CDF.

Since we have used the data to estimate the hyper-parameters $(\mu, \sigma)$, we cannot use the standard KS test to ascertain the degree of misspecification.
Instead, we must define our own KS-like test for the $p$-value using the distance between the model and data.
We calculate 100 KS distances between the model and draws from the model.
We can histogram this data to show the distribution of KS distances expected if the model is a good fit to the data; see the pink histogram in Figure \ref{fig:ks-dist-hist}.
We measure the KS distance between the average data realisation and the data-conditioned model, which is shown by a grey line in Figure \ref{fig:ks-dist-hist}.
The $p$-value equates to the fraction of the model KS distance distribution above the data KS distance.
In Figure \ref{fig:parameterised_pop:cdf}, we plot the 90\% credible bands of the parameterised model and compare against the 90\% credible range of the data.
Despite being conditioned on the same data, the misspecified model strays outside of the credible band of the data and achieves a $p$-value of only 0.01. 
Finally, in Figure \ref{fig:parameterised_pop:pp}, we show the same CDFs with the CDF of the model subtracted, and plot against the model CDF. This has the effect of emphasising the extent of the mismatch between the data and the misspecified model, while the well-specified model contains the median data draw over the entire range.

\subsection{Models with more than one dimension}
It is difficult to look for model misspecification in more than one or two dimensions. Tests like those described in this Article so far can be extended, but the tell-tale signs of misspecification (e.g., detectable structure in residuals) can be difficult to see in large-dimensional spaces. 
However, there are some tools available.
\begin{itemize}
    \item Make two-dimensional scatter plots.
    If the model does not have any parameters, it can be represented by contours while the data can be represented with two-dimensional error bars or credible intervals.
    \item If the model is subject to theoretical uncertainty (i.e., it is described by hyper-parameters), it may be necessary to draw multiple contours.
    However, the plot can be difficult to read if there are more than two variables.
    \item There is no standardised test for goodness of fit in two or more dimensions.
    Boutique tests designed to identify particular forms of misspecification can work well, but one must beware trial-factor penalties when designing the test after looking at the data.
\end{itemize}

\section{More symptoms of model misspecification}\label{other}
\subsection{Posterior stability}
In the previous Sections we describe tests to find misspecification.
However, misspecification is sometimes manifest from surprising results.
One such example is a phenomenon we refer to as ``posterior instability.''
Let us consider the posterior for some parameter (call it $\theta$) and imagine how this posterior changes as we accumulate data.
On average, we expect the posterior to narrow as we include more information.
It also tends to shift around, but only within the bounds of the previous credible regions.
It would be surprising if the addition of data produced a posterior favouring $\theta=3$ when an earlier posterior (calculated with less data) disfavoured $\theta=3$ with high credibility.

In such cases we say that the posterior is not stable to the addition of data.
This can be indicative of model misspecification.
An example from gravitational-wave astronomy is the maximum black hole mass parameter, which proved to be unstable moving from GWTC-1 \citep{o2_pop} to GWTC-2 \citep{o3a_pop}, under the assumption that the primary black hole mass distribution is a power-law with a sharp cut-off.
When the mass model was improved to allow for additional features (deviations from a power law \citep{mass}), the maximum-mass parameter stabilised.\footnote{Another example: many inferences in \cite{o3a_pop} appear to be unstable with respect to the inclusion of the extreme mass-ratio event GW190814, suggesting that the models in that work are not adequate to accommodate this event.}
For another example of posterior instability from optical astronomy, see \cite{Liu2018,SuperBayes}.

\subsection{Outliers}
Outliers are symptom of misspecification closely related to posterior stability \citep{Fishbach:2020:outliers, Essick:2021:outliers}.
An outlier is an event with a parameter value appearing inconsistent with the rest of the distribution---in the context of a particular population model.
Let us imagine that the distribution of events in our dataset (characterised by parameter $\theta$) seem well-described by a normal prior with mean zero and unit variance: ${\cal N}(\theta|\mu=0, \sigma=1$.
If we subsequently observed an event with $\theta \gtrsim 10$, this could indicate that our prior model (a normal distribution) is inadequate.

Commonly, population outliers are identified using a ``leave-one-out'' analysis method that compares an inferred distribution with and without the potentially anomalous data point; see \cite{Fishbach:2020:outliers}.
However, care must be taken to take into account trial factors\footnote{Some literature refers to the ``look elsewhere effect,'' e.g., \cite{Gross2010}.} since, in any catalog, \textit{some} event has to be the most extreme.
The statistics required for a careful leave-one-out analysis are too complicated for us to summarise here.
Instead we refer the reader to \citet{Essick:2021:outliers}, which describes a ``coarse-graining'' method to identify outliers.

\section{Discussion: living with misspecification}\label{conclusions}
Misspecified models can lead to flawed inferences.
While it is not always practical, models should, when possible, be subjected to an array of visualisations and checks for misspecification.
In an ideal world, models found to be misspecified should be improved so that they pass these checks.
In particular, one may \textit{refine} one's model in order to better capture features of the data \citep[e.g.,][]{GaoHo:2017:GalaxyDecomposition,Gabry:2017:visualisation}, although this kind of post-hoc data fitting can cause obvious biases in inferred results.
Additionally, iteratively adding complexity can cause computational costs to skyrocket, and each addition has the potential to add further misspecification.

Thus, in practice, model refinment is not always possible.
For example, as data becomes more informative (higher signal-to-noise ratio), even subtle imperfections can lead to signs of misspecification.
For example, if one attempts to fit a template to an optical image of a distant (but clearly resolved) galaxy, there will \textit{always} be non-negligible residuals because our best templates are no match for the high signal-to-noise ratio of optical astronomy data.

One option in such cases is to apply \textit{coarsening} to blur the data.
For example, one may employ a coarsened posterior, conditional on the distance between the model and the data being below some threshold, but not zero \citep[see][for a detailed description of posterior coarsening]{Miller2019}.
An alternative form of coarsening is to add a non-parameteric \textit{error term} to the revised model to account for unknown or independent influences in the data; for example, \citet{Bhatt2017ImprovedPA} use a Gaussian random field to generate multiplicative factors to the terms in their parameterised model for malaria mapping. 

Although we have described a number of ways in which models can be tested for misspecification, one must ultimately accept that \text{all} physical models are---to some degree---misspecified \citep{all-models-are-wrong}.\footnote{Put more memorably by statistician George Box as ``all models are wrong, but some are useful.''}
The question, therefore, is not whether a model is wrong, but whether it is \textit{adequate}.
If a model does a ``good enough'' job of describing a signal (that is, it is not \textit{obviously
 misspecified}), then we may still able to use it to inform us about the Universe.

\section{acknowledgments}
The authors would like to thank those who took part in the discussion during the \textit{When is a model good?} session at the 2021 OzGrav Annual Retreat, in particular Rory Smith, who co-organised and chaired the session.
We would also like to thank Ewan Cameron for his comments on the manuscript.
This work is supported through Australian Research Council (ARC) Future Fellowship FT160100112, Centre of Excellence CE170100004, and Discovery Project DP180103155.


\bibliographystyle{pasa-mnras}
\bibliography{1r_lamboo_notes}

\end{document}